\documentclass{acm_proc_article-sp}

\usepackage{stmaryrd}
\usepackage{listings}
\usepackage{color}
\usepackage{enumitem}
\usepackage{xcolor, colortbl}
\usepackage{multirow} 
\usepackage{rotating}
\usepackage{hhline}
\usepackage{amssymb}
\usepackage{algorithm2e}
\usepackage{xfrac}
\usepackage{algorithmic}
\usepackage{float}

\newcommand{\q}{\theta}  
\newcommand{\kxt}{\kappa} 

\definecolor{gray}{rgb}{0.4, 0.4, 0.4}
\definecolor{darkblue}{rgb}{0.0, 0.0, 0.6}
\definecolor{cyan}{rgb}{0.0, 0.6, 0.6}

\lstdefinelanguage{XML}
{
  morestring=[b]", 
  morestring=[s]{>}{<}, 
  morecomment=[s]{<?}{?>}, 
  stringstyle=\color{black}, 
  identifierstyle=\color{darkblue}, 
  keywordstyle=\color{cyan}, 
  morekeywords={xmlns, version, type}
}

\title{Massive Query Expansion by Exploiting \\Graph Knowledge Bases}

\numberofauthors{3}
\author{
\alignauthor
Joan Guisado-G\'amez\\
    \affaddr{DAMA-UPC}\\
    \affaddr{Universitat Polit\`ecnica de Catalunya}\\
    \email{joan@ac.upc.edu}
\alignauthor
David Dominguez-Sal
  \affaddr{Sparsity Technologies}\\
  \email{david@sparsity-technologies.com}
\and
\alignauthor
Josep-LLuis Larriba-Pey
  \affaddr{DAMA-UPC}\\
  \affaddr{Universitat Polit\`ecnica de Catalunya}\\
  \email{larri@ac.upc.edu}
}

\begin{document}

\maketitle

\begin{abstract}
Keyword based search engines have problems with term ambiguity and vocabulary
mismatch. 
In this paper, we propose a query expansion technique that
enriches queries expressed as keywords and short natural language descriptions. 
We present a new massive query expansion strategy that enriches queries 
using a knowledge base by identifying the query concepts, 
and adding relevant synonyms and semantically related terms. 
We propose two approaches: (i) lexical expansion that locates the relevant
concepts in the knowledge base; and, 
(ii) topological expansion that analyzes the network of relations among
the concepts, and suggests semantically related
terms by path and community analysis of the knowledge graph. 
We perform our expansions by using two versions of the Wikipedia as knowledge
base, concluding that the combination of both lexical and topological expansion
provides improvements of the system's precision up to more than 27\%. 

\end{abstract}

\section{Introduction}
Query expansion is the process of rewriting a query introduced by the user 
of a search engine in order to improve the retrieval performance. 
Query expansion is done under 
the assumption that the query phrase introduced by the user 
is not the most suited to
express the real intention of the user. For example, 
\textit{vocabulary mismatch} between queries and documents is one of the main 
causes of poor precision in information
retrieval systems~\cite{metzler2007similarity}. 
Poor results also arise from the \textit{topic inexperience} of the users. 
Users searching for information are often
not familiar with the vocabulary of the topic in which they search, and
hence, they may not use the most effective keywords. 
It ends up in the loss of important results due to the lack 
of detail in the query terms. 
\textit{Word ambiguity} is another 
cause of poor precision because the information retrieval system is not always
able to identify the real intention of the user. 
All these three problems become specially relevant 
in case of retrieving multimedia documents, such as videos, photos or music 
documents using their associated metadata, 
because descriptions are short and sparse~\cite{timonen2013term}.

Many query expansion techniques are based on the exploration of 
query data logs~\cite{carpineto2012survey}. Typically, 
these techniques do not introduce
noise in the keywords, but their effectiveness 
is strongly correlated to the size of the log and the success of previous
users refining the query. Besides, these techniques cannot be applied 
if there is no operating
search engine from which to collect the logs.

In the absence of a log, it has become common to use 
knowledge bases (e.g. Wikipedia, Yago, Wordnet, etc.) as 
the source of information for the query expansion. The 
entries in a knowledge base
typically have descriptions, attributes and relations between concepts. 
We mainly distinguish two approaches to exploit them in the literature:
(i) systems that identify the concepts explicit in the query in order to 
derive synonyms and close variants of the concepts~\cite{Milne2007}, 
and (ii) systems that locate relevant concepts in the 
knowledge base and include the directly linked
concepts. For example, 
in~\cite{arguello2008document} the authors present an expansion 
method that uses the anchor text of a hyperlinked document to 
enrich the queries.

In this paper, we propose a novel query expansion which uses a hybrid input
in the form of a query phrase and a context, which are a set of keywords and
a short natural language description of the query phrase, respectively. 
Our method is based on the
combination of both a lexical and topological analysis of the concepts related
to the input in 
the knowledge base. We differ from previous works because 
we are not considering the links of each article individually, 
but we are mining the global link structure of the knowledge base to find
related terms using graph mining techniques. With our technique
we are able to identify: (i) the most relevant concepts and their synonyms, 
and (ii) a set of semantically related concepts. 
Most relevant concepts provide equivalent reformulations of the query that 
reduce the vocabulary mismatch. Semantically related concepts introduce
many different terms that are likely to appear in a relevant document, which is useful
to solve the lack of topic expertise and also disambiguate the keywords.

We summarize the main contributions of this paper as:
\begin{enumerate}
\item We use a flexible hybrid search input that combines keyword 
search and short natural language descriptions, for searching in collections
with short texts. 
\item We propose a technique to construct synonym phrases that are equivalent 
to those introduced by the user.
\vspace{.3cm}
\item We propose a graph mining technique that converts
the hybrid search input into a map of paths between concepts in a knowledge 
base.
\item We propose a community detection algorithm that is able to extract
semantically related concepts and disambiguate the semantic topic
of the query. To our knowledge, this is the first query expansion proposal 
using topological information and, specifically, a community detection
algorithm.
\item We show the impact of our technique, which increases the precision 
obtained by pseudo-relevance feedback methods.
Also, we show the robustness of our proposal even in the lack of context
scenario.
\end{enumerate}

The techniques presented in the paper have been tested with two different 
knowledge bases (English Wikipedia and Simple Wikipedia). We found significant 
improvements, up to 27\% relative improvement in the precision, for both 
knowledge bases with respect to a state of the art search engine. According
to our experiments, the use of richer knowledge 
bases for extracting semantically
related terms also improves the precision of the system.

The remainder of the paper is organized as follows: in 
Section~\ref{section:overview}, we present an overview of our method. In
Section~\ref{section:lexical}, we describe the lexical based expansion of the
original query phrase. We present the topological
based expansion in Section~\ref{section:topological}. 
In Section~\ref{sec:experiments}, we present the results obtained. We review 
literature on query expansion and the usage of external data sources 
in Section~\ref{section:relatedWork}. Finally, in 
Section~\ref{section:conclusions}, we propose the future work and conclude.

\section{Massive keyword expansion\\using a Knowledge base} \label{section:overview}

 \begin{figure*}
\centering
\includegraphics[width=.9\linewidth]{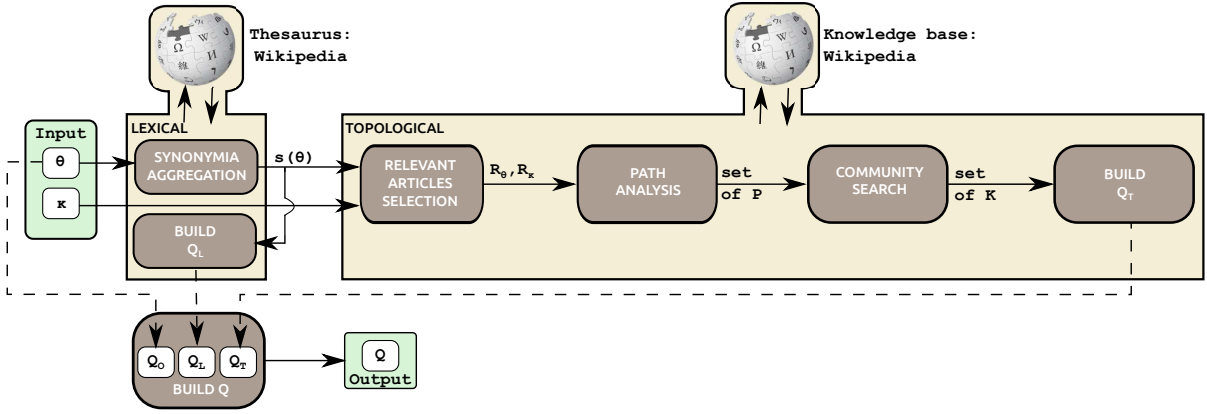}
\caption{Query extension pipeline.}
\label{fig:systemDescription}
\end{figure*}

\subsection{Knowledge base expansion}
Search sessions usually consist on a sequence of interactions with the
search interface where the user modifies an original query by adding and
removing keywords to improve search 
results~\cite{silverstein1999analysis}. 
This process is iterated by the user until the keywords introduced have enough
coverage and are not ambiguous to the search engine. Although this process has
become the \textit{de facto} search technique, this refinement and enrichment 
process is not straightforward. Humans usually introduce few words and are not
willing to use complex queries, which produces many trial and error iterations
and long search sessions.

We illustrate this difficulty with an example query from the experimental 
dataset used in this paper that looks for ``boxing match'' pictures. 
If the user is not familiar with the topic of the query, it is not 
easy to provide additional keywords that improve the search results. 
But, this information is available in a knowledge base, such as Wikipedia, 
if the system is able to identify the user search intentions. 
Note that knowledge bases do not offer always a direct mapping of keywords
and concepts. For instance, the English Wikipedia
has more than ten entries in the disambiguation page of ``Boxing'' including
sports, computer science topics, holidays, locations and songs. 
Using the connections between the articles, our system suggests 
keywords that are related to the boxing vocabulary such as ``boxing punches'' or 
``heavyweight boxing'', or boxers such as 
``Edison Pantera Miranda'' or ``Tommy Farr''.

Even if the user is an expert on the topic and can suggest the keywords, it
is also necessary to express the relevance of the keywords for the query. Some
elements are more relevant than others. Although a relevant document about 
``boxing'' may contain the word ``punch'', the former term 
should have a relevance larger than the latter because it is closer to the
information needed by the user. We also analyze the topological structure
of the relations in the knowledge base to compute distance measures and 
decide which terms are more relevant for the query.

\subsection{System overview}

\begin{table}[!t]
\begin{small}\begin{tabular}{p{.15cm}p{.15cm}||l|l|}
\cline{3-4}
\cline{3-4}
&&Symbol & Description  \\\hline
\multirow{11}{*}{\rotatebox{90} {\mbox{Fundamental}}}	&&$t$                               	& Term $t$. \\ \cline{3-4}
&\multirow{9}{*}{\rotatebox{90}{\mbox{Objects}}}    	&\multirow{2}{*}{$\rho=t_1\ t_2\ . . . \ \ t_n$}	& Phrase, i.e. an ordered \\                                                                                                   
&							&                                   	& list of terms. $|\rho|=n$ \\ \cline{3-4}
&							&$Q=[<w_1, \rho_1>, . . ., $             	& $Q$ is a query, i.e. a set \\ 
&						 	&$\ \ \ \ \ \ \ <w_q, \rho_q>]$	& of  pairs <weight, phrase>. \\ \cline{3-4}
&							&\multirow{3}{*}{$Q_X(\rho_i)=w_i$}        & Function that returns \\
&							&					&the weight associated \\
&							&					&with $\rho$ in $Q_X$.  \\\cline{3-4}
&&\multirow{3}{*}{$\varPsi$}		&Set of existing phrases \\
&							&					&in the documents. \\
&							&					&collection. \\\hline
\multirow{4}{*}{\rotatebox{90}{\mbox{Synonyms}}}&	&\multirow{2}{*}{$s(t)=\{s_1(t), s_2(t), . . . \}$}		& It is the collection of \\
&							&					&all the synonyms of $t$. \\ \cline{3-4}
&							&\multirow{2}{*}{$s(\rho)=\{s_1(\rho), s_2(\rho), . . . \}$}	& It is the collection of\\
&							&					&all the synonyms of $\rho$. \\ \cline{3-4}\hline
\multirow{9}{*}{\rotatebox{90}{\mbox{Instances}}}	&&\multirow{2}{*}{$\q$}			&Original phrase, \\
&							&					&introduced by the user. \\ \cline{3-4}
&							&\multirow{2}{*}{$\kxt$}		&Context phrase, \\
&							&					&introduced by the user. \\ \cline{3-4}
&							&$Q_O$					&Original query. \\ \cline{3-4}
&							&$Q_C$					&Context query. \\ \cline{3-4}
&							&$Q_L$					&Lexical query. \\ \cline{3-4}
&							&$Q_T$					&Topological query. \\ \cline{3-4} 
&							&$Q_E$					&Expansion query. \\ \cline{3-4} \hline
\multirow{5}{*}{\rotatebox{90}{\mbox{Wikipedia}}}	&&$a$					&Wikipedia article. \\ \cline{3-4}
&\multirow{3}{*}{\rotatebox{90}{\mbox{Objects}}}	&\multirow{2}{*}{$a^T$}			&Phrase that is the  \\
&							&					&title of article $a$. \\ \cline{3-4}
&							&\multirow{2}{*}{$a^\circlearrowleft$}	&Set of articles that are \\
&							&					&a Wikipedia redirect of $a$. \\\hline
\multirow{7}{*}{\rotatebox{90}{\mbox{Graph}}}		&&$R_{\q}$				&Relevant articles for $\q$.\\\cline{3-4}
&\multirow{5}{*}{\rotatebox{90}{\mbox{Structures}}}	&$R_{\kxt}$				&Relevant articles for $\kxt$.\\\cline{3-4}
&							&$P=a_i\rightarrow . . . \rightarrow a_j$&Path between $a_i$ and $a_j$. \\\cline{3-4}
&							&$K$					&Community of articles. \\ \cline{3-4}
&							&$\tau(P)$				& Score of $P$. \\ \cline{3-4}
&							&$\tau(K)$				& Score of $K$. \\ \cline{3-4}
&							&$h$					&Hierarchy of phrases. \\\hline\hline
\end{tabular}             \end{small}
\caption{Table of symbols.}\label{table:symbols}
\end{table}

In this paper, we follow a hybrid query input where the user provides a 
query phrase ($\q$), which is a set of keywords, and complements it 
with a context ($\kxt$), which is a
description in natural language of the query. 
Our objective is to provide an enrichment procedure that is able
to take such an input and use a knowledge base to introduce a large set of 
terms that improve the precision and the coverage of the system. 
We summarize the notation used in this article in Table~\ref{table:symbols}.

In Figure~\ref{fig:systemDescription}, we depict the 
query expansion architecture. 
First, $\q$ is processed by the lexical 
block that builds a lexical query expansion ($Q_L$). 
The lexical expansion identifies concepts of the knowledge base 
that are synonyms of the query and, after disambiguating them, includes 
them as phrases in $Q_L$. This type of expansion is aimed at finding 
terms that have the same meaning as those introduced by the user.

The second block performs a topological expansion, using both $\q$
and $\kxt$, that finds concepts that are likely to appear in
documents relevant to the query. 
This block finds concepts in the knowledge base that are relevant
to the query and the context, and connects them with paths using 
the knowledge base relations among concepts. The most relevant paths
are used as seeds of a community search algorithm that finds the closest 
concepts to those in the path. 
From these communities, the system builds the topological expansion $Q_T$.

In this paper, we are using Wikipedia as the knowledge base. We define each
Wikipedia article as a concept. In our system, each concept has an associated
phrase that corresponds to the title of the article. For the articles that are
pointed by redirect pages, the whole set of redirects is considered to describe 
the same concept. One concept is related to another if a link in 
the text of the article contains a link to the other article. This model builds
a graph structure where nodes are the Wikipedia articles that are connected
with edges that represent links. Any knowledge base that is described as a 
graph of concepts with relations, 
such as Yago~\cite{suchanek2007yago} or DBPedia, could
be used for the purpose. 

In order to illustrate the combined action of the two types of enrichment 
blocks, we will use the following
example from the experimental section:

\begin{enumerate}
 \itemsep=-.3em
 \item[]$\q$=\texttt{colored Volkswagen beetles} 
 \item[]$\kxt$=\texttt{Volkswagen beetles in any color} \\ \color{white}++\color{black}\texttt{for example, red, blue, green or yellow.}
\end{enumerate}

In this example, some of the previous difficulties appear. 
First, the term \texttt{beetles} is by itself ambiguous. In order 
to disambiguate the query, the system
should relate it to the term \texttt{Volkswagen} and locate the concept
\texttt{Volkswagen Beetle}.  
Note that the concept in our knowledge base is 
called ``Volkswagen Beetle'' and not ``Volkswagen beetles'' 
as it appears in $\q$, and thus the system should relate both concepts in order
to avoid locating a Volkswagen car with beetles. 
Second, the user introduced few words and, as we will see later, some additional
keywords can enhance the number of relevant documents retrieved.

\section{Lexical Enrichment}\label{section:lexical}
In this block, we use the information of Wikipedia as a thesaurus in order
to extract information from $\q$ from a lexical point of view. 
We compute synonyms for $\q$ based on an exact matching of the terms with 
the thesaurus, and then, we build a lexical query, $Q_L$. 

\subsection{Synonymic Aggregation with Wikipedia}
This phase computes $s(\q)$, which is the collection 
of all the synonyms of $\q$. 
First we compute the synonyms of each term, one by one, using the redirects 
of Wikipedia. 
Given a term $t$, the process retrieves (if it exists) the article $a$ from 
Wikipedia whose title $a^T$ is equal to $t$, i.e. $a^T=t$. Then, 
the synonyms of $t$ are the titles of the redirects of $a$. We only add
the redirects of articles which are formed by a single term, in other words, 
$s(t)=\{x^T\ :\ x\in a^\circlearrowleft, a^T=t, |x^T| = 1 \}$. 
Note that we look only for term-to-term synonyms in order to preserve the 
original structure of $\q$. Also, note that this module provides synonyms without 
knowledge of the language
of the search (plurals, verb forms, declensions, acronyms, etc.)
because it is based on exact matching of terms and article titles.

The synonym of the input phrase, $s(\q)$, is built as all
the combinations of term synonyms respecting the original order of the terms 
in $\q$:
$$
s(\q)= \{ s_j(\q_1) \ldots s_l(\q_n) :
1\leq j \leq |s(\q_1)|\, \ldots\, 
1\leq l \leq |s(\q_n)| \}
$$

\subsection{Lexical Query Building}
\label{sec:lexical_query_building}
We build query $Q_L$ as a set of pairs <weight, phrase> 
where the phrase in $s(\q)$ 
exists in the document collection, $\varPsi$. In other words,
phrases that are used to build $Q_L$  must match exactly
with at least one phrase in one of the documents of the document collection. We set equal weights to all
the synonyms in the lexical query:

$$Q_L=\ \left\{ \left< w, s_i(\q) \right>\ :
s_i(\q) \in \varPsi, w = \frac{1}{|Q_L|} \right\}$$

The resulting lexical query for our example is:
\[Q_L=\{<0.5,volkswagen\ beetle>, <0.5, vw\ beetle>\}\] which
is the result of replacing the term \texttt{Volkswagen} by its acronym \texttt{vw}
and the term \texttt{beetles} is replaced by its singular form \texttt{beetle}. 

If no phrases from $s(\q)$ exist in the documents collection, 
then the lexical expansion is simply $Q_L=\emptyset$. For those cases where
the introduced keywords do not perform exact matches, we rely on the expansion
performed in the topological block that analyzes the relations between concepts
of the knowledge base.

\section{Topological enrichment}\label{section:topological}
In this block, we use the topology of the knowledge base in order to extract
information about the relation among articles in order 
to find those that represent better the intent of the original query. 

\subsection{Relevant article selection}
We create two sets of relevant articles. One set is derived from
the original query phrase $R_{\q}$ and the other from the context $R_{\kxt}$. 

We obtain $R_{\q}$ from the synonyms of the 
query $s(\q)$. For each phrase in $s(\q)$, we obtain all the \textit{bigrams}.
In the example, the \textit{bigrams} of $\q$ are
[\{\texttt{colored Volkswagen}\}, \{\texttt{Volkswagen beetles}\}].
Then, we retrieve the relevant articles from Wikipedia using phrase matching 
queries, one query for each \textit{bigram}. The set of relevant articles is complemented
with the articles that contain in their title the words (\textit{unigram}) of each
phrase in $s(\q)$.

We construct $R_{\kxt}$ similarly and derive
the set of relevant articles as explained before. 

Figure~\ref{fig:relevantDocuments} shows an example of the relevant
document sets. Each circle represents a document and the connecting arrows
indicate that there is a link between them. 
In general, \mbox{$|R_{\q}|<|R_{\kxt}|$} because
$\kxt$ is usually a larger description than~$\q$. Since $\kxt$ and $\q$ are 
related it is also expected the two sets share a significant 
number of Wikipedia pages. In the example, 
pages \texttt{h} and \texttt{g} belong to $R_{\q}$ and also to $R_{\kxt}$.

\subsection{Path analysis} 
 \begin{figure}
\centering
\includegraphics[width=.75\linewidth]{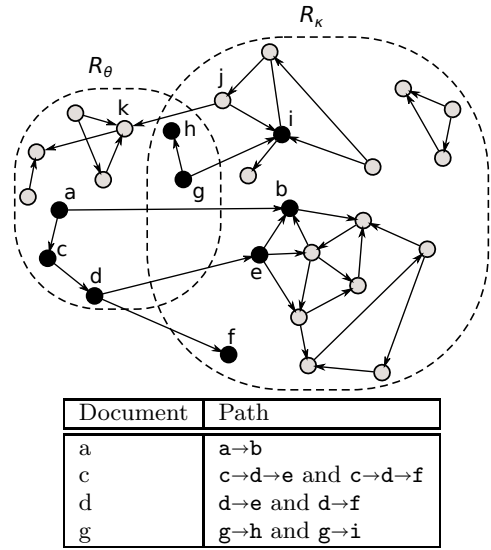}

\begin{tabular}{|l|l|}
\hline
Document & Path \\ \hline \hline
a & \texttt{a}$\shortrightarrow$\texttt{b}\\ 
c & \texttt{c}$\shortrightarrow$\texttt{d}$\shortrightarrow$\texttt{e} and \texttt{c}$\shortrightarrow$\texttt{d}$\shortrightarrow$\texttt{f} \\
d & \texttt{d}$\shortrightarrow$\texttt{e} and \texttt{d}$\shortrightarrow$\texttt{f}\\ 
g & \texttt{g}$\shortrightarrow$\texttt{h} and \texttt{g}$\shortrightarrow$\texttt{i}\\ \hline
\end{tabular}
\caption{Shortest paths from each document in $R_{\q}$.}\label{table:shortPaths}

\label{fig:relevantDocuments}
\end{figure}

Both $R_{\q}$ and $R_{\kxt}$, may contain irrelevant 
concepts that have been selected due to the ambiguity of the terms in 
$\q$ and $\kxt$. This phase builds a conceptual map
between the terms in order to disambiguate them with the aid of the 
knowledge base, and hence derive the meaning intended by the user.

For each concept in $R_{\q}$, the system computes the
shortest path that reaches a concept in $R_{\kxt}$. 
The shortest path represents the most direct way to 
connect a particular article from $R_{\q}$ to the articles in $R_{\kxt}$ and 
therefore we reduce the risk of selecting paths formed by articles that 
do not correspond to the real meaning of $\q$. 

In Figure~\ref{fig:relevantDocuments}, we depict an example where nodes 
represent articles from Wikipedia within $R_{\q}$ or $R_{\kxt}$. Darker 
nodes represent articles that are part of at least one shortest path between 
the two sets. In the table of the figure, we show all the shortest paths 
between each article of $R_{\q}$ and $R_{\kxt}$. Note that the initial 
and the final node of a path must be two different nodes, 
even if they are in the overlap of the sets. Note also that, for one 
initial node, it may exist more 
than one path. For example, there are two shortest paths that start from
\texttt{g} and reach nodes in $R_{\kxt}$: one 
goes to \texttt{h} and the other goes to \texttt{i}. 
Paths that go from $R_{\kxt}$ to $R_{\q}$, such as $j \shortrightarrow k$, 
are not considered.

Once all paths are computed, they are ranked in a descending order based on the 
score of each path. Given a path of Wikipedia pages 
$P=\texttt{a}_1\shortrightarrow 
\texttt{a}_2\shortrightarrow \dotsb \shortrightarrow \texttt{a}_s$, 
its score is 
$\tau(P)=\sfrac{\left( \sum_{i=1}^s \lambda(a_i^T, \q) + \lambda(a_i^T, \kxt) \right)}{s}$, 
where $\lambda(\rho_x, \rho_y)$ is a function that counts the number of 
terms in the intersection between $\rho_x$ and $\rho_y$. We select the 
paths with the highest score and dismiss the rest of them. 

  For the example query, the system finds 182 shortest paths using the English
Wikipedia. Among them, nine score \sfrac{3}{2} that is the top score:
\begin{itemize}
\itemsep=-.5em
     \item[] \texttt{volkswagen}$\rightarrow$\texttt{volkswagen beetle}
     \item[] \texttt{volkswagen fox}$\rightarrow$\texttt{volkswagen beetle}
     \item[] \texttt{volkswagen passat}$\rightarrow$\texttt{volkswagen beetle}
     \item[] \texttt{volkswagen type 2}$\rightarrow$\texttt{volkswagen beetle}
     \item[] \texttt{volkswagen golf}$\rightarrow$\texttt{volkswagen beetle}
     \item[] \texttt{volkswagen jetta }$\rightarrow$\texttt{volkswagen beetle}
     \item[] \texttt{volkswagen touareg}$\rightarrow$\texttt{volkswagen beetle}
     \item[] \texttt{volkswagen golf mk4 }$\rightarrow$\texttt{volkswagen beetle}
     \item[] \texttt{volkswagen beetle}$\rightarrow$\texttt{volkswagen transporter}
\end{itemize}

The first path in the list is specially relevant because it connects the 
generic concept \texttt{Volkswagen} to the most specific context 
\texttt{Volkswagen Beetle} and both are related to $\q$. The rest of the paths 
are also good because they have disambiguated the term \texttt{beetle} and 
connect specific models of \texttt{Volskwagen} with 
the model that we are interested in, \texttt{Volkswagen Beetle}.

\subsection{Community search} 
In this phase, the system enriches the previously computed paths with articles
that are closely related. The most direct solution would be to enrich the path 
with all the neighbors of the Wikipedia articles. 
However, this naive solution does not work because it introduces articles
that are loosely related to the path. Wikipedia articles have typically many
links, and many of them refer to topics that have some type of relation but
semantically are very distant. We implement a community search algorithm
to distinguish the semantically strong links from the weak ones. 

\begin{algorithm}[t]
\algsetup{linenosize=\tiny}
  \scriptsize

 \SetAlgoLined
 \KwIn{Path $P$}
 \KwOut{Community associated to a path $P$}
 $K. add(P. getArticles())$ \;
 \Repeat{currentWCC = WCC(K) }
 {
   $currentWCC \gets WCC(K)$\;
   \Repeat{bestCandidate = NULL}
   {   	
      $bestWCC \gets |K| * currentWCC $\;      
      $bestCandidate \gets NULL$\;
      $candidates \gets neighbors(K)$\;

      \ForEach{Article $c$ in $candidates$}
      {
	
	$wcc \gets WCC(K \cup c)$\;
	
	\If{$(|K|+1)*wcc>bestWCC$}
	{
	  $bestWCC\gets (|K|+1)*wcc$\;
	  $bestCandidate\gets c$\;  	  
	}
      }
      \If{$bestCandidate \neq NULL$}
      {
	$K \gets K \cup bestCandidate$\;
      }
   }
 
  \Repeat{modified=false}{
    $modified\gets false$\;

    \ForEach{Article a in K} {
	\If{WCC(a, K) < $\frac{WCC(K)}{4}$} {
	$K \gets K \setminus a$\;
	$modified \gets true$\;
	}
    }
  }
}
\caption{Average WCC maximization for $K$.} \label{alg:wcc}
\end{algorithm}

A community in a graph is a set of closely linked nodes which are similar among them but
are different from other nodes in the rest of the graph. 
We detect the communities by a process that maximizes the Weighted
Community Clustering (WCC, ~\cite{Prat2012}) of a set of nodes. The WCC($x$,$K$) 
is a metric that measures if a vertex $x$ fits in a community $K$ based on 
the number of shared transitive relations
(triangles) that $x$ has with the community. A large number of shared triangles
indicates a strong relation between the nodes~\cite{Prat2012}. The WCC of a 
community $K$, WCC($K$), is defined as the average of the WCC of the nodes in the 
community, i.e. 
WCC($K$)=$\sfrac{\sum_{\forall x \in K} WCC(x,K)}{|K|}$.

Algorithm~\ref{alg:wcc} describes our process to maximize the WCC 
of a community around a path. 
The process of creating a community around each path has two main parts: 
(i) adds vertices to the community while the sum of WCC increases; and
(ii) removes vertices while the average of WCC increases. 
In more detail, we start with a community $K$ whose vertices are the 
articles in one path. Then, we set the neighbors of the community members as
a candidate set. For each candidate $n$, we check 
whether it increases the total WCC of $K$. At the end, we add the article that 
produces a larger increase in the WCC. We keep adding vertices in $K$ while we are 
able to increase the WCC. Finally, we remove the articles that have a WCC
below \sfrac{1}{4} of the average WCC of $K$. The process is repeated until
the WCC is not improved in one iteration. The process is guaranteed to terminate
because the WCC of a community is a number between 0 and 1 and our algorithm
improves the WCC of $K$ in each iteration.

Once the communities have been created, 
we rank them in a descending order based on the score of each community, 
similarly to the selection of the paths. 
Given $K$ a community of articles, its score 
is $\tau(K)=\sum_{a \in k} \lambda(a^T, \q) + \lambda(a^T, \kxt)$. 
We select the communities with the highest score and dismiss the rest of them. 

\subsection{Topological Query Building}

In this step, the system builds $Q_T$, which scores the relevance of the 
articles in the communities already found. 

For each community $K$ found in the previous step, 
we build a hierarchy $h$, which is rooted on the terms given by the user. 
The articles are scored according to the height in the hierarchy. 
The first level of $h$ is formed by the terms in $\q$. 
The second level of the hierarchy is formed by the articles whose title
contains all the terms in $\q$. 
In other words, we include in the second level the articles $a$ such
that  $a^T \cap \q = a^T$. The $i$-th level of a hierarchy of $L$ levels
(for $2 < i \leq L$) is formed by 
the articles that have a link from an article 
in the $(i-1)$-th level. The weight of the article $a$
that is in the level $i$ of $h$, $w(a, h)$, is
computed as $ w(a, h) = \sfrac{L-i}{L-1} $. Articles that do not fit the 
previous conditions are removed from the hierarchy.
The articles placed at the top level of the hierarchy have a weight 
equal to 1 and the articles at the last level of the hierarchy have a weight 
equal to 0. Note that we are restrictive with the second level
of the hierarchy, because if no article is selected in it, then 
all articles have a weight equal to 0.

We consider all the redirects of a Wikipedia article as semantically strongly
related to the original article, and thus, we use the 
redirects as expansions, too. If the community from which the hierarchy is created
contains an article whose title is equal to a term in $\q$, then the weight
of that term within the hierarchy is calculated as the addition of the weight
of the term in the first level and the weight of the articles within the second level.

The topological expansion, $Q_T$ is constructed from 
the combination of the hierarchies. Let $H$ be the set of hierarchies, 
we average the weight of the articles across all the hierarchies where
the article is present:
$$Q_T= \left\{ <w_\rho, \rho>: \rho = a^T, 
w_\rho= \frac{ \sum_{h \in H} w(a, h)} {|H|} \right\}$$

Following the previous example, the articles selected from the English
Wikipedia, sorted by weight are: \texttt{Volkswagen}, 
\texttt{Volkswagen Beetle}, \texttt{German cars}, 
\texttt{Volkswagen group}, etc. 
The expansion is formed by 1,125 unique articles, 
each of which formed by 2.40 terms on average. 
For simplicity, we do not 
show the phrases that come from the titles of redirects.  
For example, the article \texttt{Volkswagen Beetle} has 39 redirects 
including plurals, abbreviations (\texttt{vw bug}), other phrases to 
refer the same concept (\texttt{VW Type 1}) or even frequent 
misspellings (\texttt{Volkswagon Beatle}). 
The terms shown allow us to observe 
that the topic \texttt{Volkswagen Beetle} has been disambiguated and that, 
due to topological properties, phrases such as \texttt{German cars} 
or \texttt{Volkswagen Group} are added. With smaller weights than for previous
articles, we also obtain articles 
as \texttt{Volkswagen New Beetle}, which is a newer 
version of Volkswagen Beetle, \texttt{Wolfsburg}, which is the city where 
the beetle cars were manufactured, \texttt{Baja bugs}, 
which refers to an original 
Volkswagen Beetle modified to operate off-road (open desert, sand 
dunes and beaches) or \texttt{Cal Look}, name used to refer to customized 
version of Volkswagen Beetle cars that follow a style 
oriented in California in 1969. 
Many of the terms selected correspond to terms that are not likely introduced
by the user, because although they may appear in relevant documents, 
they are not known by the user or require a research effort
to the user.

\subsection{Query Building}
In this step, we describe a process to transform the expansions found in 
the previous blocks to a query that can be computed by an information retrieval
engine that supports phrase matching and weighted terms. 
We combine the queries $Q_O, Q_L$ and $Q_T$ with the weights
$\alpha$, $\beta$ and $\gamma$ factors, respectively. 
We express the final query $Q$ 
as a structured query $\varrho$, that combines proximity
and belief operators~\cite{metzler2004combining} on $Q_E$:

 \begin{equation*}
  Q =\varrho(W, Q_E ): W=\langle \alpha, \beta, \gamma\rangle , Q_E=\langle Q_O, Q_L, Q_T\rangle 
 \label{Q_Expansion}
\end{equation*}

\begin{figure}
\centering
\includegraphics[width=1\linewidth]{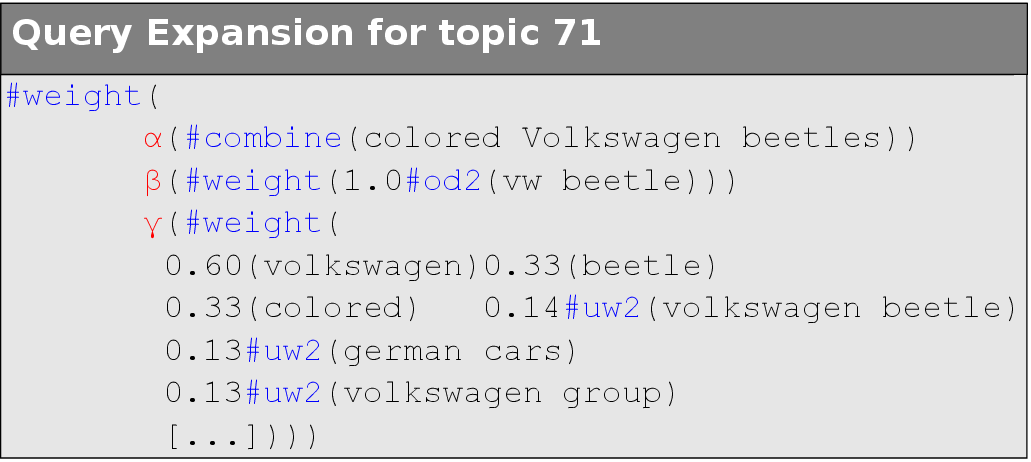}
\caption{Indri representation of the query expansion of query 71. \color{blue}\texttt{\#weight}\color{black}, \color{blue}\texttt{\#combine}\color{black}, \color{blue}\texttt{\#odN}, \color{black} and \color{blue}\texttt{\#uwN }\color{black} are part of Indri query language.}
\label{fig:queryExpansion}
\end{figure}

Figure~\ref{fig:queryExpansion} shows the structured query $Q$ for the 
example in Indri notation. In Indri, we express $Q_O$ as an unigram weighted
with $\alpha$, $Q_L$ as a set of exact matching phrases weighted with $\beta$, and
$Q_T$ as a set of unordered matching phrases weighed with $\gamma$.

If we compare $Q_O$ and $Q_E$, we observe that not 
only the concept \texttt{Volkswagen Beetle} is clearly identified, but 
some other related terms are incorporated. For example,
the concept \texttt{german cars} is obtained thanks to 
the topological structure of Wikipedia. It would be very 
difficult to derive this phrase from $\q$ without a knowledge base.

\section{Experiments}\label{sec:experiments}
\subsection{Experimental setup}

We test our query expansion method using the resources provided in the
ImageCLEF 2011 Wikipedia CLEF track. 
The test image collection contains 237,434 images downloaded from Wikipedia, 
which have short descriptions as metadata. Approximately, 60\% of these
descriptions contain texts in English. The test collection also provides
fifty queries. Each query consists of a set of keywords, a brief description
in natural language, and a set of relevant images in the test collection. 
In these experiments we use two sets of stop words. The first set is a 
collection of typical English stop words collection. The second set of stop 
words contains words that refer to the visual conditions such as colors, 
positions and shapes which our system is not able to process due to the lack
of an image processing module. 

We test our query expansion engine with two knowledge bases. The first one 
is the Simple Wikipedia, built from the dump on April 8th, 2012. It contains
112,525 articles, of which 31,564 are redirects, 
and 1,213,460 links among articles. 
The second one is extracted from the English Wikipedia, built from the dump
on July 2nd, 2012. It contains 9,483,031 articles, of which 3,3343,856 
are redirects, and 99,675,360 links among articles. 
Both Wikipedia graphs are loaded and processed using the DEX
graph database~\cite{martinez2007dex}. 

In our experiments, we used Indri\footnote{http://www.lemurproject.org/indri} as 
the search engine that processes the query and retrieves the images from the 
collection of images. Indri is a state of the art 
open-source search engine that provides 
phrase matching, term proximity, explicit term/phrase weighting and the usage
of pseudo-relevance techniques.

We set the factors $\alpha$, $\beta$ and $\gamma$ to 0.08, 0.05, 0.87, 
respectively, based on our own experience configuring the system. From now on,
consider $W=\langle 0.08,0.05,0.87 \rangle$. Note that 
the value given to the phrases obtained through the topological expansion
has one order of magnitude more importance than for the rest of factors.

\subsection{Results}

\subsubsection{Retrieval precision}
In these experiments, we measure the precision improvement obtained by the 
lexical and the topological extensions using our full query expansion system. 
We compute the precision at three
different levels. The results of the experiments
are in Table~\ref{table:results}. 

We set three baselines. 
The first baseline, $\langle Q_O \rangle$, is a traditional search
engine that relies on the small set of keywords introduced by the 
user. The second baseline, $\langle Q_O, Q_C \rangle$, 
includes the keywords and
the short description of the user. We build $Q_C$, as the set of all 
terms that appear in the context with equal weights. The third baseline, PRF, 
applies pseudo-relevance feedback, implemented by Indri, over $Q_O$.

Table~\ref{table:results} shows the results for our baselines (Base), and
the results of expanding the query phrase by using the English Wikipedia
(English) and the Simple Wikipedia (Simple) 
as our thesaurus and knowledge base. For
each Wikipedia, we show the results of the lexical expansion 
alone $\langle Q_L \rangle $, the topological expansion alone 
$\langle Q_T \rangle $, 
the lexical $\langle Q_O, Q_L \rangle$ and topological expansion 
$\langle Q_O, Q_T \rangle $ with the original keywords, and the full query expansion
as $\langle Q_O, Q_L, Q_T \rangle$. For
each Wikipedia, the best result is in bold.

\begin{table}[t]
 \begin{tabular}{m{.15cm}|l|m{.65cm}m{.35cm}|m{.65cm}m{.35cm}|m{.65cm}m{.35cm}|m{1mm}}
\cline{2-8}\cline{2-8} \cline{2-8}
&Configuration &\multicolumn{2}{|c|}{\multirow{2}{*}{P@1}}&\multicolumn{2}{|c|}{\multirow{2}{*}{P@10}}&\multicolumn{2}{|c|}{\multirow{2}{*}{P@20}}&\\
&\multicolumn{1}{c|}{$Q_E$}&&&  &&&&\\\hhline{========}
\multirow{3}{*}{\rotatebox{90} {\mbox{Base }}}&\cellcolor{gray!15}$\langle Q_O\rangle$ &\cellcolor{gray!15} 0.460&\cellcolor{gray!15} &\cellcolor{gray!15} 0.338&\cellcolor{gray!15} &\cellcolor{gray!15} 0.238&\cellcolor{gray!15} \\ 
&\cellcolor{gray!15}$ \langle Q_O, Q_C \rangle$ &\cellcolor{gray!15} 0.320&\cellcolor{gray!15} &\cellcolor{gray!15} 0.260&\cellcolor{gray!15} &\cellcolor{gray!15} 0.198&\cellcolor{gray!15} \\ 
&\cellcolor{gray!15}$ Q_O  + PRF$ &\cellcolor{gray!15} 0.400&\cellcolor{gray!15} &\cellcolor{gray!15} 0.346&\cellcolor{gray!15} &\cellcolor{gray!15} 0.283&\cellcolor{gray!15} \\ \hhline{========}

\multirow{6}{*}{\rotatebox{90} {\mbox{Simple}}}&$\langle Q_L \rangle$                                      & 0.140&                             & 0.076& & 0.055&\\[1.1ex]
&$\langle Q_T \rangle$                                      & 0.500& $^{\ \ \star}$                 & \textbf{0.362}& $^{\ \ \star\star}$  &0.278&$^{\dagger \ \star\star}$\\\cline{2-8}

&$\langle Q_O, Q_L \rangle$                         & 0.480& $^{\ \ \star}$                 & 0.358&$^{\ \ \star\star}$           & 0.255&$^{\ \ \star}$&\\[1.1ex]

&$\langle Q_O, Q_T\rangle$                         & \textbf{0.540}& $^{\ \ \star\star}$ & 0.352&$^{\ \ \star\star}$        & 0.276&$^{\dagger\ \star\star}$\\ \hhline{~-------}
&{\cellcolor{blue!15}}$\langle Q_O, Q_L, Q_T \rangle$ &{\cellcolor{blue!15}} \textbf{0.540}&\cellcolor{blue!15} $^{\ \ \star\star}$        &\cellcolor{blue!15} {0.360}&\cellcolor{blue!15}$^{\ \ \star\star}$&\cellcolor{blue!15} \textbf{0.281}&\cellcolor{blue!15}$^{\dagger\dagger\star\star}$&\\[1.1ex] \hhline{========}

\multirow{6}{*}{\rotatebox{90} {\mbox{English}}}&$\langle Q_L \rangle$                                      & 0.160&                             & 0.104& & 0.074&\\ 
&$\langle Q_T \rangle $                                      & 0.500& $^{\ \ \star\star}$                 & 0.400& $^{\dagger\ \star\star}$  &0.296&$^{\dagger\dagger\star\star}$\\ \cline{2-8}
&$\langle Q_O, Q_L \rangle$                         & 0.460& $^{\ \ \star}$                      & 0.368&$^{\dagger\ \star\star}$       & 0.259&$^{\ \ \star}$&\\[1.55ex]
&$\langle Q_O, Q_T\rangle$                         & \textbf{0.560}& $^{\dagger\ \star\star}$ & 0.394&$^{\dagger\ \star\star}$        & 0.285&$^{\dagger\ \star\star}$\\\hhline{~-------}
&\cellcolor{blue!15}$\langle Q_O, Q_L, Q_T\rangle $ &\cellcolor{blue!15} \textbf{0.560}&\cellcolor{blue!15} $^{\ \ \star\star}$        &\cellcolor{blue!15} \textbf{0.416}&\cellcolor{blue!15}$^{\dagger\dagger\star\star}$&\cellcolor{blue!15} \textbf{0.303}&\cellcolor{blue!15}$^{\dagger\dagger\star\star}$&\\[1.1ex]\hhline{========}

\end{tabular}
\caption{P@1, P@10 and P@20 with different configurations of $Q_E$. 
$^{\dagger}$/$^{\dagger\dagger}$ and $^{\star}$/$^{\star\star}$ indicate
statistically significant improvements over the $Q=\langle Q_O \rangle$ and $Q= \langle Q_O+Q_C \rangle$ 
configurations at the significance levels $0.05$/$0.01$ respectively, using 
a paired t-test. 
}
\label{table:results}
\end{table}

Our results show that the direct usage of the context reduces the precision of the
system. The reason is that the context is a short natural 
language description of the search, which is intended to be read by humans. 
In such descriptions, not all terms have equal relevance. For example, 
proper nouns are often more important than adverbs, 
and also some words, 
such as \textit{thing} or  \textit{object}, 
are used as wild cards that are not likely to appear in relevant documents. 

In our setup, pseudo-relevance feedback, $ Q_O + PRF$,
does not contribute to improve the precision with respect to $\langle Q_O \rangle$. 
PRF consists in assuming that the top results, 
obtained by running the original query, are correct. Then, those
results are used in order to extract the expansion terms and to reformulate
the original query. In the test setup, the images in the document collection
often have very short descriptions, and thus the number of coocurrent terms
retrieved by PRF techniques is sparse and not effective. This experience
justifies the need for more complex query expansion techniques 
that are not based 
on word coocurrence, in contrast to pseudo-relevance feedback.

Since the performance of $\langle Q_O, Q_C \rangle$ and $ Q_O  + PRF$  
is worse, or not significantly better, than simply using
$\langle Q_O \rangle$ , in the rest of the paper we focus our performance results 
on the comparison with the keyword only configuration.

The results show that the use of either Simple or English
Wikipedia (graph knowledge bases) for query expansion turns into an 
improvement in the performance. However, there are differences between
the usage of either. Better results are achieved for
the English one, which is larger and contains more concepts 
and links among them. That shows that our system is not only able to 
deal with large amounts of information but to benefit from it. Let us, from now on, 
focus on the use of the English Wikipedia.

The proposal that uses all the expansions described in 
the paper, $\langle Q_O, Q_L, Q_T \rangle $, achieves the best precision at all the
levels measured. 
In Table~\ref{table:results}, we show that this configuration 
obtains statistically significant improvements for all the precision levels 
with a standard confidence level 0.05. For the case of confidence level 0.01, 
we have similar results for P@10 and P@20.

According to the results, both $Q_L$ and $Q_T$, 
combined with $Q_O$, contribute
to improve the quality of the results for all the levels of precision. It is
specially remarkable the contribution of $Q_T$. The stronger boost of $Q_T$ over
$Q_L$ is explained for two reasons. (i) In our experimental environment,
we measured that the system found a $Q_L$ expansion for 32\% of the run queries.
The rest of the runs were done
with $Q_L=\emptyset$. And, (ii) $Q_T$ introduces 
many semantically related terms and
is not restricted by synonymia like $Q_L$. 
Therefore, the number of terms introduced is larger.
The P@10 and P@20 scenarios benefit more from this larger topological
expansion because they include more variants 
of the keywords, which improves the recall of the system. 

\begin{figure*}[t]
\centering
  \includegraphics[width=1\linewidth]{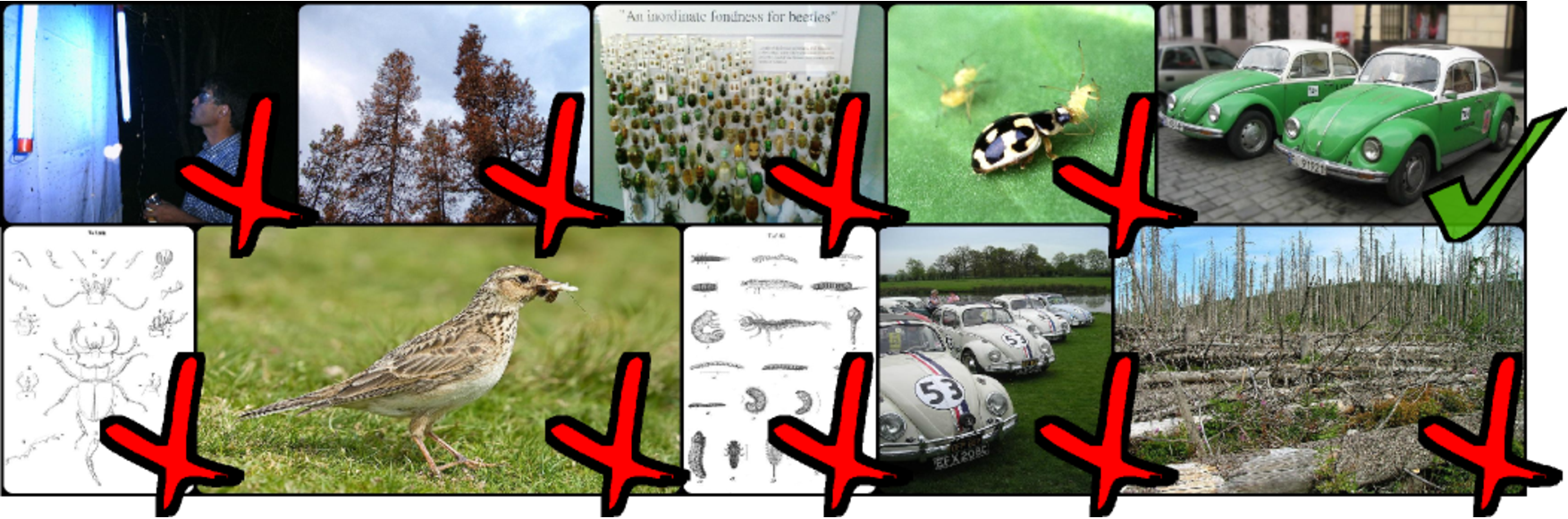}\vspace{.35cm}
  
  \includegraphics[width=1\linewidth]{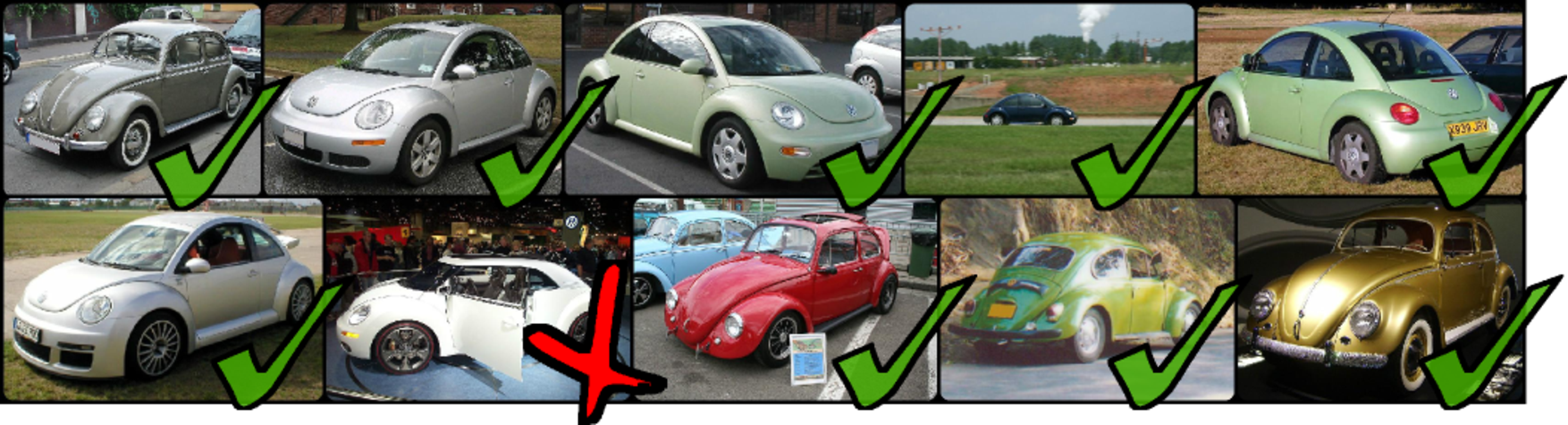}
 \caption{(color online) Top rows: Images retrieved without query 
 expansion ($Q_E=\langle Q_O \rangle $). Bottom: Images retrieved with query expansion
 ($Q_E = \langle Q_O, Q_L, Q_T \rangle $) for Query Topic 71. 
 Results are ranked from left to right, and from up to down.}
\label{fig:images}
\end{figure*}

In Figure~\ref{fig:images}, we compare the images retrieved for the baseline 
$Q_O$ (in the top rows of the figure) and the images retrieved in 
case of running $ \langle Q_O, Q_L, Q_T \rangle$ 
(in the bottom rows of the figure) for the Volkswagen Beetle example. On the one hand, the baseline
is not able to disambiguate the 
term \texttt{beetles} retrieving results mostly related to bugs due to 
the ambiguity of the term. On the other hand, after the query expansion process, 
the word has been clearly disambiguated. The disambiguation has been possible 
due to the identification of the concept \texttt{Volkswagen Beetle} and the 
addition of related terms that refer to variants of the model 
(e.g. $2^{nd}$ image corresponds to a Volkswagen New Beetle) or customized 
versions (e.g. $8^{th}$ image corresponds to a Cal Look  Volkswagen). 
Among the pictures retrieved by our proposal, the 
$7^{th}$ picture of our system is considered incorrect, 
although it is clearly a car of the desired model. The reason is that 
the query phrase ($\q$) explicitly indicates that the car must 
be colored, and the car of the picture is white. 
The current version of our system does
not include an image processing module, and thus we cannot avoid this type of
error unless the image is annotated with such information.

\subsubsection{Analysis of the expanded queries}
The construction of our expanded query relies on three queries: 
$Q_O$, $Q_L$ and $Q_T$. $Q_O$ is obtained directly from the original query
phrase ($\q$), and, thus, it needs no further analysis.

As already explained, $Q_L$ is created from the synonyms of $\q$ that exist
in $\varPsi$. Using this method, the system is able to obtain synonyms for
32\% of queries. We observed that this process is very reliable
because among all the computed $Q_L$, 100\% of them were correct
redefinitions of the query intentions. Note that the building process can 
extract phrases that are not titles or redirects in Wikipedia, as described in 
Section~\ref{sec:lexical_query_building}. We found that 70\% of the 
lexical query expansions
contain at least one phrase which is not an article in the English Wikipedia.

In Table~\ref{table:lexicalQuery}, we show some examples of 
lexical query expansions
that are not coincident with an article of Wikipedia. 
In the examples, the lexical query has disambiguated the 
real intent of the user. Results show that the lexical query is useful in 
order to identify an entity within $\q$, as for example ``\texttt{skeleton of dinosaur}''. 
Other kinds of phrases that are contained in the lexical expansion are those
that come from applying linguistic inflections to a given term (e.g. the term \texttt{flag}
has become \texttt{flags}), introducing misspellings to the given terms, using
translations to other languages for a given term (e.g. 
the terms \texttt{carnival} has been translated into german as \texttt{karneval}),
or replacing a term with its acronym (as seen 
in Section~\ref{section:lexical}, the term \texttt{Volkswagen}
has become \texttt{vw}). The lexical expansion
is relevant for our expansion method
because it introduces complex phrases which do not always correspond to
the title of the articles in Wikipedia, and hence, would not 
be included through the topological expansion.

The topological query, $Q_T$, is built from the titles of the articles within
the selected communities. 
For our test sets, the system found at least one community for 
80\% of the tested queries. Out of those communities, 85\% were communities 
semantically related with the intent of the user and 15\% were wrong. Note that
in this query set of ImageCLEF, most of the queries 
contain at least an ambiguous word.
In Table~\ref{table:communities}, we show some 
phrases of the topological queries that improve the original query phrase. We 
underline the phrases that are main articles in the English Wikipedia, and
the rest are redirects to these articles.
In order to facilitate the reading, we classify the topological expansions
in two columns: concepts that are reformulations of the query intentions, 
e.g.\ in query 80 \texttt{gray wolf} is an entity
that instances the term \texttt{wolf}; and phrases that are likely to appear 
in the same result document because of semantic relation, e.g.\ in
query 110 \texttt{William Shew} is a famous daguerrotype portrait artist. 

\vspace{.3cm}
Table~\ref{table:communities} shows also the precision for these queries with
and without expansion. We observed that the expansions provided relevant 
results to queries that initially had no relevant result, going from 0.0
to 0.9 in some cases. We also see that our query expansion
method is also effective for easy queries, which have better 
results than the average, e.g. query number 100, where the system
locates three new relevant results. 

\begin{table}[t]
\begin{tabular}{l|l|l}
ID &\multicolumn{1}{c|}{Original query phrase}&\multicolumn{1}{c}{Phrases in $Q_L$}\\
&\multicolumn{1}{c|}{($\q$)}&\\\hline\hline
\multirow{1}{*}{72}&\multirow{1}{*}{\texttt{skeleton of dinosaur}}&``\texttt{skeleton of dinosaur}''\\\hline
 \multirow{2}{*}{108}&\multirow{2}{*}{\texttt{carnival in Rio}}&``\texttt{karneval Rio}'', \\
 &&``\texttt{carnival Rio}''\\\hline
\multirow{2}{*}{118}&\multirow{2}{*}{\texttt{flag of UK}}&``\texttt{flags of uk}''\\
		     &					   &``\texttt{flag of uk}''\\\hline
\end{tabular} \caption{Phrases of $Q_L$ for ImageCLEF topics 72, 108 and 118.} \label{table:lexicalQuery}
\end{table}

\begin{table*}[t]
\begin{small}\begin{tabular}{|c||l|c||l|l|c|c|}
\hline
\multirow{2}{*}{ID}&\multirow{2}{*}{Original Query Phrase}&\multirow{2}{*}{P@10}&\multicolumn{3}{c|}{Topological expansions}&\multirow{2}{*}{P@10}\\
&&&\multicolumn{1}{c|}{Concept reformulations}&\multicolumn{1}{c|}{Semantically related}&\#Phrases&\\[1.1ex]\hline\hline
\multirow{6}{*}{80}&\multirow{6}{*}{$\q=$wolf close up}&\multirow{6}{*}{0.0}&\underline{\texttt{gray wolf}}, \texttt{wolf}, \texttt{wolve}, \texttt{timber wolves, }&\texttt{wolf evolution}, \texttt{canidae}, &\multirow{6}{*}{299}&\multirow{6}{*}{0.9}\\
&&&			\texttt{gray wolf}, \texttt{wuff}, \texttt{canis lupus, gray} & \underline{\texttt{mammal}}, {\texttt{mamalian}}, \underline{\texttt{coyote}}, &&\\
&&&			\texttt{wolves}, \texttt{grey wolf, \underline{tundra wolf}, \underline{ezo}}& \underline{\texttt{carnivora}}, \underline{\texttt{animal}}&& \\
&&&			\underline{\texttt{wolf}}, \underline{\texttt{canis dirus}}, \underline{\texttt{ethiopian wolf}}, &&&\\
&&&			\texttt{siminean jackal}, \underline{\texttt{red wolf}}, \texttt{\underline{hudson}}&&&\\
&&&			\underline{\texttt{bay wolf}}, \texttt{hudson wolf}&&&\\[1.1ex]\hline

\multirow{4}{*}{101}&&\multirow{4}{*}{0.0}&\texttt{\underline{fountain}, fountains, water}&\texttt{\underline{water}, adams ale, }&\multirow{4}{*}{67}&\multirow{4}{*}{0.6}\\
&$\q=$fountain with jet&&\texttt{fountains, wall fountain, water}&\texttt{drinking fountains, }&&\\
&of water in daylight&&			\texttt{fountain, spray fountains}&\texttt{liquid water}, &&\\
&&&			\texttt{fountain pump, waterfountain}&\texttt{water projects}&&\\
\hline
\multirow{5}{*}{110}&\multirow{5}{*}{$\q=$male color portrait}&\multirow{6}{*}{0.0}&\texttt{\underline{portrait}, portraitist, portaiture}, &\texttt{\underline{william shew}, \underline{yevgeniy}}&\multirow{5}{*}{52}&\multirow{5}{*}{0.5}                           \\
&&&			\texttt{ritratto, celebrity portrait, }&\texttt{\underline{fiks}} &&\\
&&&			\texttt{\underline{portrait painting}, portrait-painter, }&&& \\
&&&			\texttt{\underline{self-portrait}, autoritratto, }&&&\\
&&&			\texttt{autoportrait, \underline{portrait photography}}&&&\\[1.1ex]\hline

\multirow{7}{*}{100}&\multirow{7}{*}{$\q=$brown bear}&\multirow{7}{*}{0.6}&\texttt{\underline{bear}, ursine, arctos, ursidae, }&\texttt{\underline{asian black bear}}, & \multirow{7}{*}{327} & \multirow{7}{*}{0.9}\\
&&&		 	\texttt{ursoidea, bears, \underline{brown bear}}, &\texttt{\underline{tibetan blue bear}, }&&\\
&&&			\texttt{mountain bear, wild bear, broan}&\texttt{\underline{black bear}, \underline{ursus}}&&\\
&&&			\texttt{bear, american brown bears, \underline{eurasian}}&\texttt{\underline{minimus}, \underline{caniformia}, }&&\\
&&&			\texttt{\underline{brown bear}, european brown bear, }&\texttt{\underline{mammal}}, \texttt{\underline{animal}, \underline{asia}, }&&\\
&&&			\texttt{\underline{brown bear}, caucasian bear}, &\texttt{\underline{north america}}&&\\
&&&			\texttt{\underline{syrian himalayan brown bear}}&&&\\[1.1ex]\hline
\end{tabular}\end{small}
\caption{Most relevant phrases in the topological query built from queries 80, 101, 110, 100. Phrases that are underlined come from articles in Wikipedia, phrases that are not underlined are come from their redirects.}\label{table:communities}
\end{table*}

Regarding the 15\% of queries which have non semantically 
related communities, the query expansion only reduces the quality 
for one of them. The reason is that these queries were difficult
in their original formulation, and were below the average precision 
and originally returned few relevant results with Indri. 
For example, query number 
79: \texttt{heart shaped} is specially difficult because it describes an image
abstraction with text, therefore, it has 
an strong visual component that our system is not 
able to deal with. The topological extension of this query, roots in the 
biological concept of heart and, consequently, it contains related phrases
such as such as: \texttt{human heart, 
cardiac, circulatory system}, etc.

\subsubsection{Contextless query expansion}
%
%
%
In this section we discuss the robustness of our method in an scenario where
context is missing. Although our system is able to
use the short natural language descriptions, some search engines lack
a context field. 
We set all the contexts of the query set as the original query: $\kxt$ = $\q$.
This implies that
the paths described in Section~\ref{section:topological} are done within a 
single set.

Table~\ref{table:resultsContextless} shows the precision of our method after
the modifications. In this scenario, we observe that our method is still able to 
achieve an improvement of 17\% in the best situation.
Comparing Table~\ref{table:results} and Table~\ref{table:resultsContextless},
we observe that the context is specially useful in case of using the English
Wikipedia, which is larger than the Simple Wikipedia. This implies that the
usage of a query description is specially relevant in case of using a large 
knowledge base, where input terms can be matched to more articles.


Our technique still works in the absence of a natural language context 
provided by the user. However, using a short description in natural language
for the query allows the system to achieve a better performance.

\begin{table}[t]
 \begin{tabular}{m{.87cm}|l|m{.65cm}m{.20cm}|m{.65cm}m{.20cm}|m{.65cm}m{.20cm}|m{1mm}}
\cline{2-8}\cline{2-8} \cline{2-8}
&Configuration &\multicolumn{2}{|c|}{\multirow{2}{*}{P@1}}&\multicolumn{2}{|c|}{\multirow{2}{*}{P@10}}&\multicolumn{2}{|c|}{\multirow{2}{*}{P@20}}&\\
&\multicolumn{1}{c|}{$Q_E$}&&&  &&&&\\\hhline{========}

\multirow{1}{*}{{\mbox{Base}}}&\cellcolor{gray!15}\begin{small}$\langle Q_O \rangle$                                                        \end{small} &\cellcolor{gray!15} 0.460&\cellcolor{gray!15} &\cellcolor{gray!15} 0.338&\cellcolor{gray!15} &\cellcolor{gray!15} 0.238&\cellcolor{gray!15} \\ \hhline{========}

\multirow{1}{*}{{\mbox{Simple}}}&{\cellcolor{blue!15}}\begin{small}$\langle Q_O, Q_L, Q_T \rangle $                                                                              \end{small} &{\cellcolor{blue!15}} {0.540}&\cellcolor{blue!15}        &\cellcolor{blue!15} {0.360}&\cellcolor{blue!15}&\cellcolor{blue!15} {0.271}&\cellcolor{blue!15}$^{\dagger}$&\\[1.1ex] \hhline{========}


\end{tabular}
\caption{P@1, P@10 and P@20 with different configurations of $Q_E$. 
$^{\dagger}$ indicates statistically significant improvements over the $Q_E=Q_O$ 
configuration at the significance levels of $0.05$ using a paired t-test.\vspace{-.5cm}}
\label{table:resultsContextless}
\end{table}

\vspace{2cm}

\section{Related work}\label{section:relatedWork}
Query expansion techniques can be classified 
according to the methods applied in order to obtain the expansion features,
into several families~\cite{carpineto2012survey}: 
linguistic analysis, query specific, query-log analysis, and linked
data techniques.

Query expansion through \textbf{linguistic analysis} aims to extract the 
expansion features through the languages properties such as morphological, 
lexical, semantic or syntactic. These techniques expand each word of the 
original query independently of the fully query. Consequently, these query 
expansion techniques suffer from word sense ambiguity~\cite{carpineto2012survey}. Most traditional 
techniques of this family are based on stemming~\cite{Paice1994}. 
However, more exhaustive evaluations of stemming techniques reveal that is 
not always a good choice and sometimes effects negatively the precision of the 
expanded query~\cite{Gauch1999}. More recent techniques that take into account 
morphological variants have been proposed~\cite{Moreau2007}. 
Ontology analysis is also used in order to obtain the expansion 
features from a linguistic perspective~\cite{bhogal2007review}~\cite{navigli2003analysis}. 
Ontologies range from general (e.g. WordNet~\cite{miller1995wordnet}) to 
domain-specific (e.g. in the medical~\cite{diaz2009query} and legal 
domains~\cite{diaz2009query}). Query expansion through ontology analysis 
suffers from vocabulary mismatch between the original query terms and the 
concepts in the ontology. 

\textbf{Query specific} techniques exploit the set of top ranked documents
to iteratively apply relevance feedback techniques. The process can be improved
by applying clustering techniques along
the iterations\cite{DBLP:journals/ipm/ChangOK06}.

Query expansion through \textbf{query-log analysis} intends to exploit the 
information in the logs, such as the click activity of the users. 
Logs contain two 
different types of valuable information for the query expansion problem. 
The first one is the transformations that user apply over the original query. 
In~\cite{Boldi2008}~and~\cite{Song2012} the authors induce a graph 
representation of query transformations. This graph is then used to expand the 
queries. The second valuable information in the logs is the relation between 
queries and selected documents. In~\cite{Cui2002} the authors extract 
probabilistic correlations between query terms and document terms by analyzing 
query logs. These correlations are then used to select expansion terms for the 
expanded query. Query-log analysis has proved to be very useful to obtain 
high-quality query expansion. However it is an unfeasible technique when the 
system lacks large logs, as in the system that we are proposing here. 

\textbf{Linked data techniques} take advantage of web and data corpora, 
similarly to the technique presented in this paper.
Web data analysis, such as anchor texts (also known as hyperlinks 
or links), is also used as a source of information for query expansion. 
In~\cite{Eiron2003} the authors show that anchor texts are similar to real 
queries regarding to term distribution and length. 
In~\cite{dang2010query}, propose to use anchor text in the web to derive a simulated
query log in a web test collection. 
Those
techniques are orthogonal to those presented in the paper, and hence, we could
add anchor analysis to provide more synonyms of the terms detected by our
graph mining algorithm.

Corpus analysis is also used in query 
expansion techniques. The idea behind
this family is to identify relevant information for the query. Wikipedia
has become a frequent large corpus of information. 
For example, Egozi et al.~\cite{egozi2011}
present an interesting technique for query rewriting 
based on explicit semantic analysis, where they postprocess queries obtained
from pseudo-relevance feedback using a knowledge base. This
technique depends on the quality of the pseudo-relevance feedback expansion,
which is very poor in our document collection, 
as seen in Section~\ref{sec:experiments}.
In~\cite{hu2009understanding}
the authors propose a system in order to disambiguate the user's intent linking
each query to a concept within a map of concepts.
The authors propose to preprocess Wikipedia pages in order to build the map of
concepts, in contrasts to our proposal which uses directly the content and structure of
Wikipedia.
Koru~\cite{Milne2007} is a search interface that exploits the content of the 
Wikipedia in order to derive a thesaurus. The basic idea is to use the 
articles from Wikipedia as building phases for the thesaurus, and its skeleton 
structure of hyperlinks to determine which phases are needed and how they 
should fit together. Note that in this proposal they need to derive a thesaurus 
from the content in Wikipedia. Our proposal differs from this approach because 
we use directly the information in Wikipedia and do not need to construct a 
thesaurus according to the its content. 
In~\cite{arguello2008document}, the authors propose a query 
expansion method for blog recommendation. Their method is based on the analysis
of links. The anchor text of most important twenty links is used to expand the 
query which results in a significant improvement in terms of precision. 
Such an approach could be used in our work to rate the importance of the links, and
then, include the strength of connections in our community detection 
algorithm. 
In contrast to previous works, in this paper we do not limit our analysis 
to direct links of the articles, but on the full topology of Wikipedia in order
to identify articles that are related to the query.

\section{Conclusions and Future Work}
\label{section:conclusions}

In this paper, we have presented two novel contributions to the query expansion 
area. First, we have shown that the exploitation of contexts and knowledge
bases leads us to smarter search engines that are better
in terms of performance than traditional keyword based systems. However,
contexts cannot be exploited naively. And thus, second, 
we have presented a new query expansion technique which uses 
the Wikipedia in order to disambiguate the query according to its context and
complete it with semantically related terms to reduce word mismatch and
topic inexperience. In this work, Wikipedia is not only used for 
disambiguating but it is also used in order to suggest new phrases that will be 
added to the original query from a lexical and a topological point of view. The 
combination of both types of phrases achieves 
significant better results. Our experiments
also show a correlation between the precision of the system and the quality 
of the knowledge base, which suggests that the advances in creating
more complete or customized knowledge bases will provide better search engines.

We have obtained significant improvements compared 
to the baseline using the path analysis of the Wikipedia and the proposed 
community search algorithm. However, we believe that some aspects 
of the knowledge base can be further exploited. \texttt{\texttt{}}
For example, our current system does not differentiate 
among links. Some links appear in 
sections that are more relevant, or may have an special meaning such
as indicating where a person is born or temporal information of an event.
We believe that such knowledge could be introduced into the community detection
procedures to improve the quality of the system. 

\vspace{.3cm}
\bibliographystyle{abbrv}
\begin{small}\bibliography{wsdm2013}
\vspace{1cm}
\end{small}

\end{document}